\documentclass[runningheads, 10pt]{llncs}

\usepackage[T1]{fontenc}
\usepackage[utf8]{inputenc}

\usepackage{amssymb, amsmath}
\usepackage{xcolor}
\usepackage{wrapfig}

\usepackage{tabularray}
\UseTblrLibrary{booktabs}
\usepackage{braket}
\usepackage{algorithm}
\usepackage[noend]{algpseudocode}

\usepackage[style=lncs,natbib=true,maxcitenames=1, backend=bibtex]{biblatex}

\usepackage[unicode]{hyperref}
\usepackage[capitalise,noabbrev]{cleveref}

\usepackage{tikz,pgfplots}
\usepgfplotslibrary{groupplots}
\usetikzlibrary{calc}
\pgfplotsset{compat=1.18}

\usepackage{xstring}
\usepackage{xparse}

\usepackage{mathtools}
\usepackage{xcolor}
\usepackage{moresize}
\usepackage{graphicx}
\usepackage[inline]{enumitem}
\usepackage{tikzpeople}
\usepackage{booktabs}
\usepackage{tabularx}
\usepackage{xspace}
\usepackage[many]{tcolorbox}

\usepackage{subcaption}

\usepackage{todonotes}
\usepackage{pifont}

\usepackage[adversary,sets,lambda,operators,probability,logic,landau]{cryptocode}

\usetikzlibrary{decorations,shapes,arrows,backgrounds,patterns,matrix,
	fit,calc,shadows,plotmarks,topaths,chains,positioning}

\usepackage{orcidlink}
\renewcommand{\orcidID}[1]{\orcidlink{#1}}

\crefname{construction}{construction}{constructions}
\crefname{protocol}{proto.}{proto.}
\crefname{definition}{def.}{def.}
\crefname{theorem}{thm.}{thm.}
\crefname{proposition}{prop.}{prop.}
\crefname{figure}{fig.}{fig.}
\crefname{figure*}{fig.}{fig.}
\crefname{section}{sec.}{sec.}
\crefname{equation}{eq.}{eq.}
\crefname{assumption}{assumption}{assumptions}
\crefname{lemma}{lem.}{lemmas}

\newcommand{\msf}[1]{\ensuremath{\mathsf{#1}}\xspace}
\newcommand{\mcal}[1]{\ensuremath{\mathcal{#1}}\xspace}

\newcommand{\alg}[1]{\msf{#1}}

\definecolor{cadmiumgreen}{rgb}{0.0, 0.42, 0.24}

\makeatletter
\renewcommand*\env@matrix[1][*\c@MaxMatrixCols c]{%
	\hskip -\arraycolsep
	\let\@ifnextchar\new@ifnextchar
	\array{#1}}
\makeatother

\NewDocumentCommand{\oneinputalg}{R(){Alg}m}{%
	\IfBlankTF{#2}{%
		\ensuremath{{#1}}%
	}{%
		\ensuremath{{#1}\left(#2\right)}%
	}\xspace
}

\NewDocumentCommand{\twoinputalg}{R(){Alg}mom}{%
	\IfBlankTF{#4}{%
		\ensuremath{{#1}}%
	}{%
		\IfValueTF{#3}{%
			\ensuremath{{#1}\left(#3,#4\right)}%
		}{%
			\ensuremath{{#1}\left(#2,#4\right)}%
		}%
	}\xspace
}

\NewDocumentCommand{\threeinputalg}{R(){Alg}momom}{%
	\IfBlankTF{#6}{%
		\ensuremath{#1}%
	}{%
		\IfValueTF{#5}{%
			\IfValueTF{#3}{%
				\ensuremath{{#1}\left(#3,#5,#6\right)}%
			}{
				\ensuremath{{#1}\left(#2,#5,#6\right)}%
			}%
		}{%
			\IfValueTF{#3}{%
				\ensuremath{{#1}\left(#3,#4,#6\right)}%
			}{%
				\ensuremath{{#1}\left(#2,#4,#6\right)}%
			}%
		}%
	}\xspace
}

\NewDocumentCommand{\fourinputalg}{R(){Alg}momomom}{%
	\IfBlankTF{#8}{%
		\ensuremath{#1}%
	}{%
		\IfValueTF{#7}{%
			\IfValueTF{#5}{%
				\IfValueTF{#3}{%
					\ensuremath{{#1}(#3,#5,#7,#8)}%
				}{
					\ensuremath{{#1}(#2,#5,#7,#8)}%
				}%
			}{%
				\IfValueTF{#3}{%
					\ensuremath{{#1}(#3,#4,#7,#8)}%
				}{%
					\ensuremath{{#1}(#2,#4,#7,#8)}%
				}%
			}%
		}{%
			\IfValueTF{#5}{%
				\IfValueTF{#3}{%
					\ensuremath{{#1}(#3,#5,#6,#8)}%
				}{
					\ensuremath{{#1}(#2,#5,#6,#8)}%
				}%
			}{%
				\IfValueTF{#3}{%
					\ensuremath{{#1}(#3,#4,#6,#8)}%
				}{%
					\ensuremath{{#1}(#2,#4,#6,#8)}%
				}%
			}%
		}%
	}\xspace
}

\NewDocumentCommand{\optindex}{om}{%
	\IfValueTF{#1}{%
		\ensuremath{{#2}_{#1}}%
	}{%
		\ensuremath{{#2}}%
	}\xspace
}

\newtcolorbox{conceptbox}[2][]{
	breakable,
	vfill before first=false,
	segmentation at break=false,
	size=fbox,
	colback=w,
	title={\scriptsize\textbf{{#2}}},
	left=2pt,
	right=2pt,
	top=3pt,
	bottom=1pt,
	boxrule=1pt,
	coltitle=b,
	colupper=b,
	pad at break=5pt,
	toprule at break=4pt,
	bottomrule at break=0.75pt,
	colframe=#1,
	before upper*={\setlength{\baselineskip}{0.75em}\setlength{\parskip}{0em}}
}

\definecolor{w}{HTML}{fafafa}
\definecolor{g}{HTML}{bbbbbb}
\definecolor{b}{HTML}{000000}
\definecolor{lightgrey}{HTML}{cccccc}

\NewDocumentEnvironment{highlightbox}{ O{gray} O{} }% 
  {%
    \begin{tcolorbox}[
      breakable,                     	% Allow box to break across pages
      colframe=#1,                   	% Use the first optional argument as border color
      colback=#1!10,                 	% Use the first optional argument as background color (10% opacity)
      boxrule=0.4pt,                 	% Thin border width
      arc=2mm,                       	% Rounded corners
      fontupper=\small,       				% Text in \footnotesize
      width=\dimexpr\linewidth\relax, % Adjust width to align with itemize content
      left=1mm,                      	% Left padding
      right=1mm,                     	% Right padding
      top=1mm,                       	% Top padding
      bottom=1mm,                    	% Bottom padding
      segmentation at break=false,
      toprule at break=1pt,
      bottomrule at break=0.75pt,
      #2
    ]
  }
  {%
    \end{tcolorbox}
  }

\usetikzlibrary{arrows.meta,graphs,quotes,decorations,intersections,calc}

\NewDocumentCommand{\FFq}{o}{\optindex[#1]{\FF}}
\NewDocumentCommand{\ZZp}{o}{\optindex[#1]{\ZZ}}

\makeatletter
\newcommand{\tpmod}[1]{\mkern 3mu({\operator@font mod}\mkern 6mu#1)}
\makeatother

\Crefname{theorem}{Thm.}{Thms.}
\Crefname{definition}{Def.}{Defs.}
\Crefname{proposition}{Prop.}{Props.}
\Crefname{corollary}{Cor.}{Cors.}

\NewDocumentCommand{\pk}{o}{\optindex[#1]{\msf{pk}}}
\NewDocumentCommand{\sk}{o}{\optindex[#1]{\msf{sk}}}
\NewDocumentCommand{\ek}{o}{\optindex[#1]{\msf{ek}}}
\NewDocumentCommand{\vk}{o}{\optindex[#1]{\msf{vk}}}
\NewDocumentCommand{\skey}{o}{\optindex[#1]{\msf{k}}}

\NewDocumentCommand{\data}{o}{\optindex[#1]{\msf{d}}}
\NewDocumentCommand{\aux}{o}{\optindex[#1]{\msf{\tau}}}
\NewDocumentCommand{\msg}{o}{\optindex[#1]{\msf{m}}}
\NewDocumentCommand{\ctxt}{o}{\optindex[#1]{\msf{c}}}
\NewDocumentCommand{\secr}{o}{\optindex[#1]{\msf{s}}}
\NewDocumentCommand{\fout}{o}{\optindex[#1]{\msf{y}}}

\NewDocumentCommand{\lagrange}{o}{\optindex[#1]{\lambda}}

\NewDocumentCommand{\tx}{o}{\optindex[#1]{\msf{tx}}}
\NewDocumentCommand{\rx}{o}{\optindex[#1]{\msf{rx}}}

\ProvideDocumentCommand{\sig}{o}{\optindex[#1]{\msf{\sigma}}}
\NewDocumentCommand{\wit}{o}{\optindex[#1]{\omega}}
\NewDocumentCommand{\zkproof}{o}{\optindex[#1]{\pi}}

\NewDocumentCommand{\rules}{o}{\optindex[#1]{\rho}}
\NewDocumentCommand{\auxin}{o}{\optindex[#1]{\msf{a}}}

\ProvideDocumentCommand{\share}{o}{\optindex[#1]{\msf{s}}}
\NewDocumentCommand{\secshare}{o}{\optindex[#1]{\msf{z}}}

\NewDocumentCommand{\hashf}{om}{\oneinputalg(\optindex[#1]{\alg{H}}){#2}}
\NewDocumentCommand{\shaf}{om}{\oneinputalg(\optindex[#1]{\alg{SHA256}}){#2}}
\NewDocumentCommand{\keccakf}{om}{\oneinputalg(\optindex[#1]{\alg{Keccak256}}){#2}}

\NewDocumentCommand{\invfunc}{om}{\oneinputalg(\optindex[#1]{\alg{Inv}}){#2}}
\NewDocumentCommand{\ffunc}{om}{\oneinputalg(\optindex[#1]{\alg{f}}){#2}}
\NewDocumentCommand{\gfunc}{om}{\oneinputalg(\optindex[#1]{\alg{g}}){#2}}
\NewDocumentCommand{\iffunc}{om}{\oneinputalg(\optindex[#1]{\alg{f}^{-1}}){#2}}
\NewDocumentCommand{\igfunc}{om}{\oneinputalg(\optindex[#1]{\alg{g}^{-1}}){#2}}
\NewDocumentCommand{\fphi}{om}{\oneinputalg(\optindex[#1]{\alg{\phi}}){#2}}

\NewDocumentCommand{\testf}{om}{\oneinputalg(\optindex[#1]{\alg{T}}){#2}}
\NewDocumentCommand{\testin}{om}{\oneinputalg(\optindex[#1]{\alg{T}_{i:}}){#2}}
\NewDocumentCommand{\testout}{om}{\oneinputalg(\optindex[#1]{\alg{T}_{o:}}){#2}}

\NewDocumentCommand{\fee}		{o}	{\optindex[#1]{\alg{F}}}
\NewDocumentCommand{\feein}		{o}	{\optindex[#1]{\alg{F}_{i:}}}
\NewDocumentCommand{\feeout}	{o}	{\optindex[#1]{\alg{F}_{o:}}}
\NewDocumentCommand{\feeval}	{o}	{\optindex[#1]{\alg{F}_{\validator:}}}

\NewDocumentCommand{\complex}{om}{\oneinputalg(\optindex[#1]{\mcal{O}}){#2}}

\NewDocumentCommand{\sign}{om}{\twoinputalg(\alg{Sign}){\sk[]}[#1]{#2}}
\NewDocumentCommand{\sigver}{od()m}
	{\threeinputalg(\alg{Ver}){\pk[]}[#1]{\sig}[#2]{#3}}

\NewDocumentCommand{\encrypt}{om}{\twoinputalg(\alg{Enc}){\pk[]}[#1]{#2}}
\NewDocumentCommand{\decrypt}{om}{\twoinputalg(\alg{Dec}){\sk[]}[#1]{#2}}
\NewDocumentCommand{\heval}{om}{\twoinputalg(\alg{Eval}){\ek[]}[#1]{#2}}

\NewDocumentCommand{\nizkprove}{om}{\twoinputalg(\alg{ZK.Prove}){\wit[]}[#1]{#2}}
\NewDocumentCommand{\nizkver}{om}{\twoinputalg(\alg{ZK.Ver}){\zkproof[]}[#1]{#2}}

\NewDocumentCommand{\inflaf}{om}{\oneinputalg(\optindex[#1]{I}){#2}}
\NewDocumentCommand{\inflafp}{om}{\oneinputalg(\optindex[#1]{I^{\prime}}){#2}}
\NewDocumentCommand{\yieldf}{om}{\oneinputalg(\optindex[#1]{Y}){#2}}
\providecommand{\target}{\alg{t}}
\providecommand{\tot}{\alg{tot}}
\providecommand{\equi}{\alg{eq}}
\providecommand{\base}{\alg{b}}
\providecommand{\ideal}{\alg{id}}
\providecommand{\maxinfl}{\alg{M}}
\NewDocumentCommand{\decay}{o}{\optindex[#1]{\alg{d}}}

\NewDocumentCommand{\partyX}{om}{\ensuremath{\optindex[#1]{\mcal{#2}}}\xspace}

\NewDocumentCommand{\sender}			{o}{\partyX[#1]{S}}
\NewDocumentCommand{\receiver}		{o}{\partyX[#1]{R}}
\NewDocumentCommand{\party}				{o}{\partyX[#1]{X}}

\NewDocumentCommand{\publisher}		{o}{\partyX[#1]{P}}
\NewDocumentCommand{\guardian}		{o}{\partyX[#1]{G}}
\NewDocumentCommand{\validator}		{o}{\partyX[#1]{V}}
\NewDocumentCommand{\calculator}	{o}{\partyX[#1]{C}}
\NewDocumentCommand{\revoker}			{o}{\partyX[#1]{T}}
\NewDocumentCommand{\danode}			{o}{\partyX[#1]{D}}
\NewDocumentCommand{\retriever}		{o}{\partyX[#1]{R}}
\NewDocumentCommand{\actor}				{o}{\partyX[#1]{A}}

\NewDocumentCommand{\business}			{o}{\partyX[#1]{B}}
\NewDocumentCommand{\developer}		{o}{\partyX[#1]{D}}
\NewDocumentCommand{\user}		{o}{\partyX[#1]{U}}

\NewDocumentCommand{\advX}{om}{\optindex[#1]{\mcal{#2}}}
\NewDocumentCommand{\advA}{o}{\advX[#1]{A}}

\providecommand{\polkadot}{\alg{Polkadot}}

\NewDocumentCommand{\DAA}{o}{\optindex[#1]{\alg{DAA}}}
\NewDocumentCommand{\DAC}{o}{\optindex[#1]{\alg{DAC}}}

\newcommand{\PoS}{\msf{PoS}}

\NewDocumentEnvironment{HL-example}{}
	{\begin{highlightbox}[gray][ignore nobreak,fontupper=\small]}{\end{highlightbox}}

\NewDocumentEnvironment{HL-question}{}
	{\begin{highlightbox}[gray][ignore nobreak,fontupper=\small]}{\end{highlightbox}}

\NewDocumentEnvironment{HL-answer}{}
	{\begin{highlightbox}[yellow][ignore nobreak,fontupper=\small]}{\end{highlightbox}}

\title{Stabilizing the Staking Rate, Dynamically Distributed Inflation and Delay
	Induced Oscillations}
\titlerunning{On Stabilizing the Staking Rate}

 \author{
 	Carlo Brunetta\inst{1}\orcidID{0000-0001-9363-7585}
 	\and
 	Amit Chaudhary\inst{2}
 	\and\\
 	Stefano Galatolo\inst{3}\orcidID{0000-0003-3934-5412}
 	\and
 	Massimiliano Sala\inst{4}\orcidID{0000-0002-7266-5146}
 }
 \authorrunning{Brunetta, Chaudhari, Galatolo, Sala}

 \institute{
 	Independent Researcher, France, \email{brunocarletta@gmail.com}
 	\and
 	University of Warwick, UK, \email{amit.chaudhary.3@warwick.ac.uk}
 	\and
 	University of Pisa, Italy \email{stefano.galatolo@unipi.it}   
 	\and
 	University of Trento, Italy \email{massimiliano.sala@unitn.it}
 }

\bibliography{lib/finalbiblio}

\begin{document}
	
	\maketitle
    
	% To allow compilation of the file
% !TeX root = ./../main.tex

\begin{abstract}
	Dynamically distributed inflation
	is a common mechanism used to guide a blockchain's staking rate towards a
	desired equilibrium between
	network security and token liquidity.
	However, the high sensitivity of the annual percentage yield to changes in the staking
	rate, coupled with the inherent feedback delays in staker responses, can induce undesirable
	oscillations around this equilibrium.
	
	This paper investigates this instability phenomenon.
	We analyze the dynamics of inflation-based reward systems and propose a novel distribution
	model designed to stabilize the staking rate.
	Our solution effectively dampens oscillations, stabilizing the yield within a target staking range.
	
	\keywords{
		Tokenomics \and
		Dynamically Distributed Inflation \and
		Delay Induced Oscillation
	}
\end{abstract}

% Plaintext Abstract

% Dynamically distributed inflation is a common mechanism used to guide a blockchain's staking rate towards a desired equilibrium between network security and token liquidity.

% However, the high sensitivity of the annual percentage yield to changes in the staking rate, coupled with the inherent feedback delays in staker responses, can induce undesirable oscillations around this equilibrium.

% This paper investigates this instability phenomenon.

% We analyze the dynamics of inflation-based reward systems and propose a novel distribution model designed to stabilize the staking rate.

% Our solution effectively dampens oscillations, stabilizing the yield within a target staking range.

	% Content
	% To allow compilation of the file
% !TeX root = ./../main.tex

\section{Introduction}\label{sec:introduction}

A fundamental economic design challenge for Proof-of-Stake (\PoS)~\cite{AFT:BSAAMM19,RFS:Saleh20}
protocols is maintaining
the staking rate within a specific optimal
range~\cite{SSRNEJ:JohRivSal21,SSRNEJ:KogFanVis23,BRA:LeoRov24}.
To reach this target, it is essential to find a trade-off between network security and token liquidity.
A staking rate below the desired threshold reduces the crypto-economic security of the
consensus mechanism by lowering the cost of a malicious attack.
Conversely, an excessively high staking rate impairs the protocol's economic vitality by
diminishing the circulating supply, which can lead to low on-chain liquidity, increased price
volatility, and a constrained transactional ecosystem.
Therefore, designing incentive mechanisms to guide the staking rate into some desired target
corridor is a primary objective of a protocol's monetary policy.
To achieve this balance, many modern blockchains implement a dynamic inflation mechanism where
rewards distributed to stakers are a function of the current staking rate,  maximizing
the individual yield as this rate approaches the desired value~\cite{AMR:ClaSch25,MDPIFT:RipGup24}.
This approach of adjusting rewards based on the current state of the network is conceptually analogous to the ``flexible inflation targeting'' rules studied in modern monetary economics, where central banks manage a trade-off between inflation stability and output-gap stabilization \cite{BOOK:GiaWoo04}.

As an example of dynamical-inflation protocols in the PoS context,  \polkadot have implemented reward
mechanisms that tie stakers' rewards to token inflation, following a
non-linear curve designed to maximize the inflation distributed to stakers at the target
\textit{staking rate}~\cite{Polkadot,SSNR:ConHeTan2022} (see \Cref{fig:inflation_curve}).
While an approach of this kind is effective in creating a strong incentive towards an
equilibrium point, its architecture presents {\em dynamical vulnerabilities.}
Indeed, the yield curve typically exhibits a very steep gradient (high sensitivity)
around the target point meaning that a small deviation in the \textit{staking rate} causes a
sharp change in the yield, stimulating a strong corrective action in the economic agents.
When coupled with high sensitivity, the staker reaction to these yield changes becomes a natural \emph{delayed feedback} that, as it will be discussed in this work, can create the conditions for the emergence of oscillations in the staking rate.

Economic literature has identified how the combination of high reactivity and decisional
lags can generate cycles and instability in dynamical systems~\cite{QJE:Ezekiel38,RES:Kaldor34}.
In the context of staking, this delay is both structural 
and behavioural.
Recent literature indeed highlights that the design of staking mechanisms involves inherent delays, such as lock-up periods to ensure validator quality \cite{ARXIV:OdeMarHaf24} and managed exit queues to preserve protocol security \cite{ARXIV:NeuPaiRes24}. 
Further empirical evidence confirms this dynamics, showing that the reward rate in one period
is a significant predictor of an increase in the \emph{staking rate} in the next period.
Moreover, empirical analysis indicates that the combined structural and behavioral delay in
staker responses is of the order of one week \cite{XXX:ConHeTan25}.
As we will see in the following, this delayed reaction can cause the system to continuously over-correct, oscillating around
the desired equilibrium rather than converging to it stably.
The stability of the staking rate is further complicated by the presence of alternative on-chain yield opportunities, such as lending protocols. As rational agents reallocate their capital between staking and lending in pursuit of the highest return, they can induce significant volatility in the total amount staked, potentially compromising network security, a phenomenon modeled and analyzed in \cite{ARXIV:Chitra20}.

In this work, however we focus on the study of the effect of the sensitivity and the structural delay, showing that this is sufficient to generate oscillations in the staking rate.
We study this phenomenon by modeling the aggregate behavior of economic agents as a single "mean" staker, who adjusts their staking position in proportion to the difference between the current yield and a target yield level.
While this target yield is held constant in our model, we acknowledge it is influenced by external
factors, such as alternative market investments.
In \Cref{sec:model}, through numerical simulations, we compare the evolution of the staking rate across
three distinct inflation distribution schemes, showing how oscillations can be dampened by the design of an
appropriate inflation curve.
In addition to these simulations, by a suitable linearization of the system around its equilibrium
points, in \Cref{sec:anal} we present analytical results that quantitatively connect the emergence
of oscillations to the derivative of the inflation curve at the optimal staking point.

Given that the delays are a necessary structural feature and, as we have seen, a primary driver of the oscillatory dynamics, the above analysis is strongly motivated by the need to manage instability by focusing on the remaining control variable: the design of the dynamic inflation curve.

To assess the impact of the reward curve's design on system stability, we compare three inflation-distribution schemes.
As benchmarks, we use a constant inflation model where rewards are independent of the staking rate
and a Polkadot-style model characterized by a high-sensitivity yield peak.
Against these, we test a particular inflation curve, which is engineered to reduce oscillations
by replacing the yield peak with a robust ``stability corridor''. 
The core objective of this design is to nullify the yield's sensitivity within a predefined
target range, thus removing the main driver of instability, while preserving strong corrective
incentives outside this corridor.

Both numerical and analytical results confirm that the combined effect of the sensitivity of
the stakers, delay and high derivative of the yield at the equilibrium points generate
persistent oscillations, which can be dumped by reducing the derivative of the inflation at
the equilibrium point.

Our approach of analyzing staking rate stability through the lens of control theory aligns with recent research applying formal methods and dynamical systems to blockchain economies. This literature has explored the impact of various factors on protocol stability, such as the competitive pressure from on-chain lending yields \cite{ARXIV:Chitra20}, the complex emergent behaviors in lending markets \cite{ARXIV:BarLip25}, or the adaptive management of token supplies for infrastructure networks \cite{ARXIV:ASOVC22}.

	% To allow compilation of the file
% !TeX root = ./../main.tex

\section{Oscillations and Dynamical Distribution of the Inflation}\label{sec:stakerate}

In this section, we develop a dynamic model to quantitatively analyze the impact of the reward
curve's design on the stability of the staking rate.
The model simulates the aggregate behavior of stakers as a discrete-time process, where the
change in staking is driven by the difference between the current yield and a target yield
level $\yieldf[\target]{}$, incorporating a reaction delay consistent with empirical observations.
We utilize this framework to compare the system's evolution under three distinct inflation
distribution schemes:
a baseline constant-inflation model,
a Polkadot-inspired model featuring a high-sensitivity yield peak,
and a third mechanism based on a stability corridor designed to stabilize the convergence.
Through numerical simulations, we will observe the time trajectory of the staking rate for each
scenario, and the related  presence, amplitude, of damping of oscillations, thereby assessing
the effectiveness of each design in promoting a stable equilibrium.

\begin{figure}[!ht]
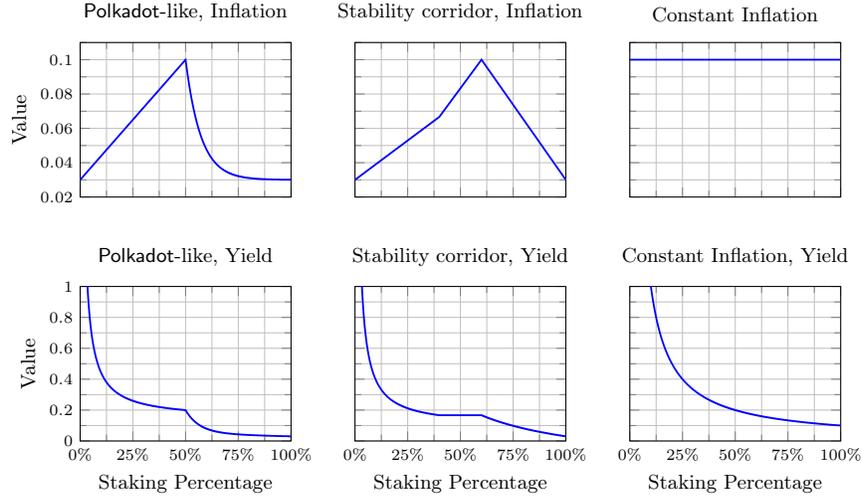

	\centering
	\scalebox{0.85}{
	\begin{tikzpicture}

	\definecolor{cred}{rgb}{1,0.8,0.8} 
	\definecolor{cyellow}{rgb}{1,1,0.8}
	\definecolor{cgreen}{rgb}{0.8,1,0.8}
	
	\def\bwidth{0.8}
	\def\bheight{1.0}
	\def\ymax{1}

	\begin{groupplot}[
			group style={
					group size=3 by 2,
					vertical sep=14mm, 
					ylabels at=edge left,
					yticklabels at=edge left,
				},
			width=.4*\textwidth,
			height=40mm,
			grid=both,
			minor tick num=1,
			xlabel={Staking Percentage},
			ylabel={Value},
			xmin = 0, xmax = 1,
			ymin = 0, ymax = \ymax,
			xtick distance = 0.25,
			ytick distance = 0.20,
			tick label style={font=\footnotesize},
			xlabel style={font=\footnotesize},
			ylabel style={font=\footnotesize},
			scaled y ticks=false,               % Disable scaling that might trigger scientific notation
			yticklabel style={
				font=\scriptsize,
				/pgf/number format/fixed,       % Use fixed point notation
				% /pgf/number format/precision=2, % Number of decimal places to show
			},
			xticklabel = {
				\pgfmathparse{\tick*100}\pgfmathprintnumber{\pgfmathresult}\%
			},
			xticklabel style={
				font=\scriptsize,
			},
		]
		
		% -------------------------------------------------------
		% -------------------------------------------------------
		
		\nextgroupplot[
			title={\polkadot -like, Inflation },
			ymax = 0.11,
			ymin = 0.02,
			ytick distance = 0.02,
			xticklabels={},
			xlabel={},
		]
		
		\begin{pgfonlayer}{main}	
			\input{fig/data/inflation-polkadot}
		\end{pgfonlayer}

		% -------------------------------------------------------
		
		\nextgroupplot[
			title={Stability corridor, Inflation },
			ymax = 0.11,
			ymin = 0.02,
			ytick distance = 0.02,
			xticklabels={},
			xlabel={},
		]
		
		% Automatically Generated 
		\begin{pgfonlayer}{main}
			\input{fig/data/inflation-mindcrypt}
		\end{pgfonlayer}
		
		% -------------------------------------------------------
		
		\nextgroupplot[
			title={Constant Inflation},
			ymax = 0.11,
			ymin = 0.02,
			ytick distance = 0.02,
			xticklabels={},
			xlabel={},
		]
		
		% Automatically Generated 
		\begin{pgfonlayer}{main}
			\input{fig/data/inflation-constant}
		\end{pgfonlayer}

		% -------------------------------------------------------
		% -------------------------------------------------------
		
		\nextgroupplot[
			title={\polkadot -like, Yield},
			xlabel={Staking Percentage},
		]
		
		\begin{pgfonlayer}{main}
			\input{fig/data/yield-polkadot}
		\end{pgfonlayer}

		% -------------------------------------------------------
		
		\nextgroupplot[
			title={Stability corridor, Yield},
			xlabel={Staking Percentage},
		]
		
		\begin{pgfonlayer}{main}
			\input{fig/data/yield-mindcrypt}
		\end{pgfonlayer}
		
		% -------------------------------------------------------
		
		\nextgroupplot[
			title={Constant Inflation, Yield},
			xlabel={Staking Percentage},
		]
		
		\begin{pgfonlayer}{main}
			\input{fig/data/yield-constant}
		\end{pgfonlayer}

	\end{groupplot}

	\end{tikzpicture}}
	\caption{
    The total inflation curve $\inflaf{\sigma}$ as a function of the staking rate $\sigma$
    for the three examples described in the text.}
	\label{fig:inflation_curve}
\end{figure}

We now describe the three dynamical inflation distribution mechanisms we compare.
We will denote by $\sigma$ the current staking rate, i.e. the fraction of the total supply being
staked and by $\inflaf{\sigma}$ the distributed total inflation, depending on $\sigma$.
\Cref{fig:inflation_curve} graphically illustrates the three distributed-inflation functions and
the relative Annual Percentage Yield (APY).

% !Tex Root = ./../main.tex

\paragraph*{\textbf{Constant Reward.}}
In this reward scheme, $\inflaf{\sigma}$ is constant.
We set it at $0.1$ ($10\% $) in our study, and this will be the maximum inflation rate also
provided by the other dynamic-reward mechanisms studied here.
Keeping constant the total distributed inflation, the individual yield per staker,
$\yieldf{\sigma}$, is inversely proportional to the staking rate, $\sigma \in [0,1]$, as the
same amount of rewards is divided among a larger staking base according to
$\yieldf{\sigma} = \frac{\inflaf[\tot]{\sigma}}{\sigma}$.

In a simple situation like this one, the equilibrium staking rate $\sigma_{\equi}$ (where the
individual yield equals the target yield $Y_{\target}$) is then 
$\sigma_{\equi}  = \frac{\inflaf[\tot]{}}{\yieldf[\target]{}}$.
There can be oscillations around this fixed point, depending on the sensitivity of the behavior
of the stakers, as it will be discussed in the next sections.

\paragraph*{\textbf{Polkadot-like Reward.}}
A well-known example of a dynamic-reward mechanism is the \polkadot inflation model, which
utilizes a piecewise function to steer the staking rate towards an ideal equilibrium.
The model's behaviour is governed by several key parameters:
\begin{itemize}[label={$\circ$}]
  \item $\sigma_{\ideal}$: the target staking rate (in our study it will be set to $0.5$).
  \item $\inflaf[0]{}$: the minimum annual inflation rate (which we set to $0.03$).
  \item $\inflaf[\maxinfl]{}$: the maximum inflation rate, reached when $\sigma = \sigma_{\ideal}$
  	(we set to $0.10$).
  \item $\decay$: a decay parameter controlling how quickly inflation decreases for staking rates
  	above the ideal target (which we set to $0.05$).
\end{itemize}
The total inflation function is formally defined by the following equation:
\begin{equation}
	\inflaf{\sigma} = 
	\begin{cases} 
		\inflaf[0]{} + \left( \frac{\sigma}{\sigma_{\ideal}} \right)
			\cdot (\inflaf[\maxinfl]{} - \inflaf[0]{}) &
			\text{if } 0 < \sigma \le \sigma_{\ideal} \\
		\inflaf[0]{} + (\inflaf[\maxinfl]{} - \inflaf[0]{})
			\cdot 2^{ \left( \frac{\sigma_{\ideal} - \sigma}{\decay} \right) } &
			\text{if } \sigma > \sigma_{\ideal}.
	\end{cases}\label{eq:polkadot_inflation}
\end{equation}

The function consists of two distinct regimes.
For staking rates at or below the target, inflation increases linearly, creating a progressively
stronger incentive to stake.
Conversely, for staking rates above the ideal target, inflation decays exponentially, which strongly
disincentivizes further staking to protect the token's liquidity and utility within the ecosystem.
However as we will see, this rapid decrease of the reward, coupled with the delayed action of the
stakers easily stimulate oscillations in the staking rate, we will discuss in next section.

\paragraph*{\textbf{Stability Corridor Reward.}}
This inflation distribution system is based on a piecewise-affine total inflation function,
defined by three distinct regimes.

We assume a target range for the \emph{staking rate} between $40\%$ and $60\%$.
The inflation function will be constructed in a way that in this range the yield is constant.
As before, we set the inflation to range between a minimum of $3\%$ to a maximum of $10 \%$ and
to vary continuously with three affine branches.
These three affine branches will be constituted by an \textbf{incentivization phase} for
$0\leq \sigma <0.4$, where the inflation grows linearly and the associated yield per token is very
high promoting the investment. 
Then, for $0.4\leq \sigma \leq 0.6$, we have a \textbf{stability corridor} in this interval,
inflation continues to grow linearly, but in a way that the individual yield is constant.
This is the core of our stabilization mechanism.
By nullifying the change in yield, the sensitivity that, combined with delay, generates
instability is eliminated.
Agents have no incentive to alter their positions in response to fluctuations of the
\emph{staking rate} within this zone, thus promoting a stable equilibrium.
Then, we have for $0.6<\sigma \leq 1.0$ a \textbf{disincentivization phase} where to prevent
excessive token illiquidity, the inflation decreases linearly from its $10\%$
peak back to the base value $\inflaf[\base]{} = 3\%$ when $\sigma =1.0$  and the yield also
decreases similarly, discouraging further staking.

To satisfy the above requirements, $\inflaf{\sigma}$ is defined as follows:
\begin{equation}
\inflaf{\sigma} =
	\begin{cases}
		0.03 + 0.025\cdot \sigma  			& \text{if } 0 		\leq 	\sigma < 		0.4 \\ 
		0.1 \cdot \sigma  							& \text{if } 0.4 	\leq 	\sigma \leq 0.6 \\ 
		0.105 - 0.075\cdot \sigma  			& \text{if } 0.6 	< 		\sigma \leq 1.0
	\end{cases}\label{eq:infl}
\end{equation}

	% To allow compilation of the file
% !TeX root = ./../main.tex

\section{A Dynamical Model for the Staking Rate Evolution}\label{sec:model}

The preceding discussion suggests that the sensitivity of the yield to variations in the
staking rate, combined with the delayed response of economic agents, may promote oscillatory
behaviour around the equilibrium staking rate.
In this section, we develop a model to illustrate, at least qualitatively, the emergence of
this phenomenon and to assess the plausibility that this dynamic is an important driver of instability.

We simulate the aggregate behaviour of stakers under each of the three dynamic-reward mechanisms
described previously.
We model the ``mean'' agent as deciding to increase or decrease their staked position in proportion
to the difference between the current yield and a constant target yield.
This decision is then implemented with a seven-day delay, a lag consistent with empirical
observations~\cite{SSNR:ConHeTan2022}.
The model's dynamics are simulated and analyzed for the three inflation distribution curves
outlined in the previous section, using a target yield parameter of $16.6\%$.

\vspace{1mm}

We formalize the above ideas, measuring the time in days.
The staking rate at time $n$ will be denoted by $\sigma_n \in [0,1]$.
The value of $\sigma_n$ is determined by the value on the previous day, $\sigma_{n-1}$,
plus an adjustment term which is proportional to the difference between the yield observed
seven days prior, $\yieldf{\sigma_{n-7}}$, and a constant target yield, $\yieldf[\target]{}$.
The proportionality constant, $b > 0$, represents the sensitivity of the ``mean'' staker to
yield deviations.
We hence get the following recurrence relation, where the result is clamped to the
valid range $[0,1]$:
\begin{equation}
    \sigma_n = \max\left( \vphantom{\Big()}  0,
    	\min\left( \vphantom{\big()} 1, \sigma_{n-1} + 
    		\left(\vphantom{\big()} \yieldf{\sigma_{n-7}} - \yieldf[\target]{} \right) \cdot b
    	\right)
    \right)
    \label{eq:staking_dynamic}
\end{equation}

{
The yield at day $n-7$, denoted as $\yieldf{\sigma_{n-7}}$, is not an independent variable but it is itself a function of the staking rate at that time. It is determined by the active inflation distribution curve, $\inflaf{\sigma}$, according to the relation:
\begin{equation}
   \yieldf{\sigma_{n-7}} = \frac{I(\sigma_{n-7})}{\sigma_{n-7}}.
    \label{eq:yield_function}
\end{equation}
}
It is important to note that, since the yield function $\yieldf{}$ is non-linear, the model
described in \Cref{eq:staking_dynamic} is a \emph{non-linear dynamical system with
delay}.
The model's global dynamics can hence be complex and difficult to study rigorously. 

\begin{figure}[!ht]
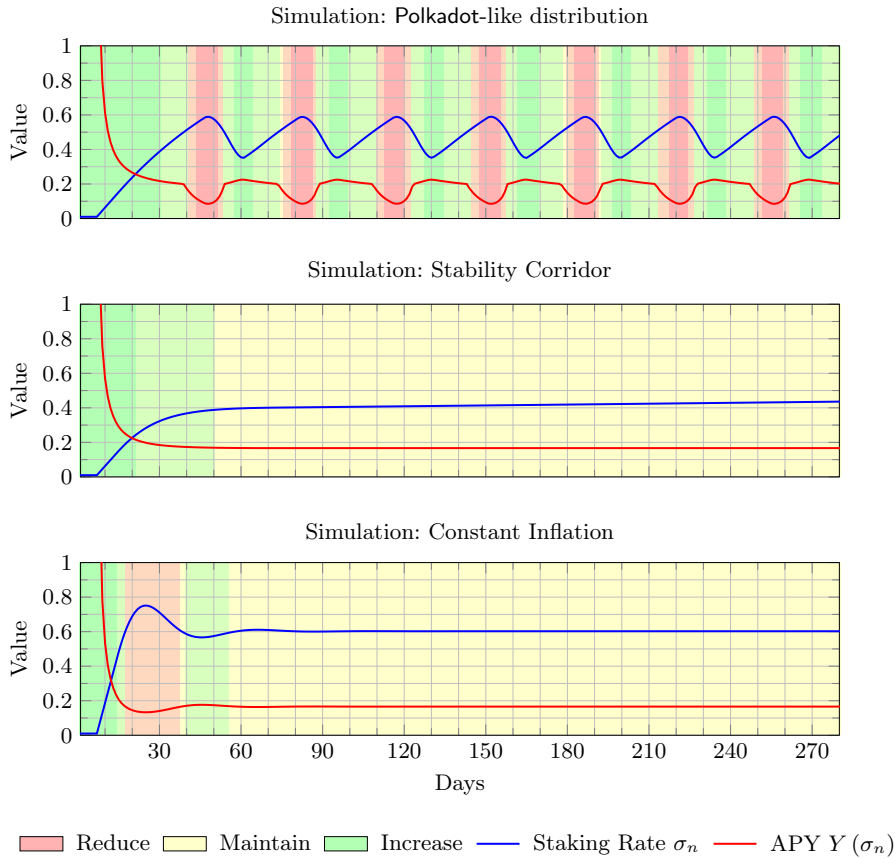

	\centering
	
	\definecolor{cred}{rgb}{1.0,0.7,0.7}
	\definecolor{credorange}{rgb}{1.0,0.85,0.75}
	\definecolor{cyellow}{rgb}{1.0,1.0,0.8}
	\definecolor{cyellowgreen}{rgb}{0.85,1.0,0.75}
	\definecolor{cgreen}{rgb}{0.7,1.0,0.7}
	
	\def\bwidth{0.8}
	\def\bheight{1.0}
	\def\ymax{1.0}
	
	\def\lsupM{0.05}
	\def\lsupB{0.003}
	\def\linfB{-\lsupB}
	\def\linfM{-\lsupM}

	\scalebox{0.95}{
		\begin{tikzpicture}

			\begin{groupplot}[
					group style={
							group size=1 by 3,
							vertical sep=12mm, 
							ylabels at=edge left,
							yticklabels at=edge left,
						},
					width=\textwidth,
					height=40mm,
					xlabel={Days},
					ymin = 0, ymax = \ymax,
					ylabel={Value},
					grid=both,
					minor tick num=2,
					minor y tick num=1,
					xmin = 1, xmax = 280,
					xtick distance = 30,
					ytick distance = 0.2,
					tick label style={font=\small},
					xlabel style={font=\small},
					ylabel style={font=\small},  
				]
				
				% --- 
				\nextgroupplot[
					title={Simulation: \polkadot-like distribution},
					xticklabels={}, 
					xlabel={},
				]
				
				\pgfdeclarelayer{background}
				\pgfdeclarelayer{main}
				\pgfdeclarelayer{grid}

				\pgfsetlayers{background,main,grid} % Activate the layers

				% Automatically Generated 
				\begin{pgfonlayer}{main}
						\input{fig/data/sim-polkadot-stake}
						\input{fig/data/sim-polkadot-apy}
				\end{pgfonlayer}

				\begin{pgfonlayer}{background}
					% Define the data points for filling the background
					\input{fig/data/sim-polkadot-action}
					
					% Plot the bars
					\foreach \datapoint in \mydatapoints {

						\StrBetween{\datapoint}{(}{,}[\xvalue]
						\StrBetween{\datapoint}{,}{)}[\yvalue]
						
						% Define the color based on yvalue
						
						\ifdim\yvalue pt < \linfM pt
							\addplot+[ybar, fill=cred, no markers, draw=none, bar width=\bwidth, bar shift=0pt]
								coordinates { (\xvalue , \bheight) };
						\else\ifdim\yvalue pt < \linfB pt
							\addplot+[ybar, fill=credorange, no markers, draw=none, bar width=\bwidth, bar shift=0pt]
								coordinates { (\xvalue , \bheight) };
						\else\ifdim\yvalue pt < \lsupB pt
							\addplot+[ybar, fill=cyellow, no markers, draw=none, bar width=\bwidth, bar shift=0pt]
								coordinates { (\xvalue , \bheight) };
						\else\ifdim\yvalue pt < \lsupM pt
							\addplot+[ybar, fill=cyellowgreen, no markers, draw=none, bar width=\bwidth, bar shift=0pt]
								coordinates { (\xvalue , \bheight) };
						\else
							\addplot+[ybar, fill=cgreen, no markers, draw=none, bar width=\bwidth, bar shift=0pt]
								coordinates { (\xvalue , \bheight) };
						\fi\fi\fi\fi
					};
				\end{pgfonlayer}

				\nextgroupplot[
					title={Simulation: Stability Corridor},
					xticklabels={}, 
					xlabel={},
				]

				\begin{pgfonlayer}{main}
					\input{fig/data/sim-mindcrypt-stake}
					\input{fig/data/sim-mindcrypt-apy}
				\end{pgfonlayer}

				\begin{pgfonlayer}{background}
					% Define the data points for filling the background
					\input{fig/data/sim-mindcrypt-action}

					% Plot the bars
					\foreach \datapoint in \mydatapoints {

						\StrBetween{\datapoint}{(}{,}[\xvalue]
						\StrBetween{\datapoint}{,}{)}[\yvalue]
						
						% Define the color based on yvalue
						
						\ifdim\yvalue pt < \linfM pt
							\addplot+[ybar, fill=cred, no markers, draw=none, bar width=\bwidth, bar shift=0pt]
								coordinates { (\xvalue , \bheight) };
						\else\ifdim\yvalue pt < \linfB pt
							\addplot+[ybar, fill=credorange, no markers, draw=none, bar width=\bwidth, bar shift=0pt]
								coordinates { (\xvalue , \bheight) };
						\else\ifdim\yvalue pt < \lsupB pt
							\addplot+[ybar, fill=cyellow, no markers, draw=none, bar width=\bwidth, bar shift=0pt]
								coordinates { (\xvalue , \bheight) };
						\else\ifdim\yvalue pt < \lsupM pt
							\addplot+[ybar, fill=cyellowgreen, no markers, draw=none, bar width=\bwidth, bar shift=0pt]
								coordinates { (\xvalue , \bheight) };
						\else
							\addplot+[ybar, fill=cgreen, no markers, draw=none, bar width=\bwidth, bar shift=0pt]
								coordinates { (\xvalue , \bheight) };
						\fi\fi\fi\fi
						
					};
				\end{pgfonlayer}
				
				\nextgroupplot[
					title={Simulation: Constant Inflation}
				]

				\begin{pgfonlayer}{main}
					\input{fig/data/sim-constant-stake}
					\input{fig/data/sim-constant-apy}
				\end{pgfonlayer}

				\begin{pgfonlayer}{background}
					% Define the data points for filling the background
					\input{fig/data/sim-constant-action}

					% Plot the bars
					\foreach \datapoint in \mydatapoints {

						\StrBetween{\datapoint}{(}{,}[\xvalue]
						\StrBetween{\datapoint}{,}{)}[\yvalue]
						
						% Choose color based on y-value range and use appropriate addplot command
						\ifdim\yvalue pt < \linfM pt
							\addplot+[ybar, fill=cred, no markers, draw=none, bar width=\bwidth, bar shift=0pt]
								coordinates { (\xvalue , \bheight) };
						\else\ifdim\yvalue pt < \linfB pt
							\addplot+[ybar, fill=credorange, no markers, draw=none, bar width=\bwidth, bar shift=0pt]
								coordinates { (\xvalue , \bheight) };
						\else\ifdim\yvalue pt < \lsupB pt
							\addplot+[ybar, fill=cyellow, no markers, draw=none, bar width=\bwidth, bar shift=0pt]
								coordinates { (\xvalue , \bheight) };
						\else\ifdim\yvalue pt < \lsupM pt
							\addplot+[ybar, fill=cyellowgreen, no markers, draw=none, bar width=\bwidth, bar shift=0pt]
								coordinates { (\xvalue , \bheight) };
						\else
							\addplot+[ybar, fill=cgreen, no markers, draw=none, bar width=\bwidth, bar shift=0pt]
								coordinates { (\xvalue , \bheight) };
						\fi\fi\fi\fi

					};
					
				\end{pgfonlayer}
				
			\end{groupplot}

			% --- Legend
				
			\path (group c1r3.south) -- (group c1r3.south) coordinate[pos=0.5] (legendpos);
			\node[below=10mm of legendpos] {
				\begin{tikzpicture}
					\begin{axis}[
							hide axis,
							xmin=0, xmax=1,
							ymin=0, ymax=1,
							legend columns=10,
							legend style={draw=none, column sep=0.8ex},
						]
						% Automatically Generated 
						\addlegendimage{area legend, fill=cred}
						\addlegendentry{Reduce}
						\addlegendimage{area legend, fill=cyellow}
						\addlegendentry{Maintain}
						\addlegendimage{area legend, fill=cgreen}
						\addlegendentry{Increase}
						\addplot+[no markers, thick] coordinates {(0,5)};\addlegendentry{Staking Rate $\sigma_n$};
						\addplot+[no markers, thick] coordinates {(0,5)};\addlegendentry{APY $\yieldf{\sigma_n}$};
						% \addplot+[draw=none] coordinates {(0,5)};\addlegendentry{Stacking Action};
						% Automatically Generated 
					\end{axis}	
			\end{tikzpicture}
			};

		\end{tikzpicture}
	}
	\caption{The dynamics of the staking rate with the three proposed inflation distribution curves.
		In blue we plot the staking rate as a function of time, in red the relative yield. 
		The actions taken by the staker (increase, maintain or reduce staking) are plotted as
		background colour.}
	\label{fig:simul:polkadot}\label{fig:simulation}
\end{figure}

Our first method of analysis will therefore be numerical simulation.
\Cref{fig:simulation} illustrates the dynamics of the staking rate resulting from the simulation
of our model under the three inflation-distribution schemes previously introduced:
the \polkadot-like model, the stability corridor model and the constant inflation model.
All simulations share similar base parameters: they start from an initial condition of $\sigma_0 = 1\%$, while the sensitivity parameter is set to $b=0.6$ and the target yield is $\yieldf[\target]{} = 16.6\%$.
The results  show that the \polkadot-like model produces strong and persistent oscillations
around the optimal rate.
In contrast, the stability corridor model effectively dampens this oscillatory behavior, converging
smoothly to a stable equilibrium within the target range.
The constant inflation model exhibits an intermediate behavior, showing some initial oscillations
that are eventually damped as the system settles into its final state.

% !Tex Root = ./../main.tex

\vspace{3mm}

An alternative, simpler model of the same kind could be considered where
the adjustment term is not scaled by the staking rate. The dynamic would then be described
by the equation $\sigma_n = \sigma_{n-1} + b \cdot (\yieldf{\sigma_{n-7}}) - \yieldf[\target]{})$.
This represents a system where the absolute change in the staking rate is proportional to the
absolute yield gap, implying that the pool of capital ready to react to yield deviations is
constant, regardless of the current number of stakers. Although more direct, we believe the
model presented in the main text is more realistic, particularly at low staking rates, where new
investors may be reluctant to invest in the system. In any case, the two models present similar qualitative behaviour.

	% To allow compilation of the file
% !TeX root = ./../main.tex

\section{Stability and Oscillations near the Equilibrium Points}\label{sec:anal}

To analytically investigate the emergence of oscillations, we study the local stability of the
model's equilibrium points.
The model's adjustment term can be rewritten, revealing an alternative economic interpretation.

Denoting by $ \Delta \sigma_n \coloneqq \sigma_n -\sigma_{n-1}$ the change in the staking, we get:
\begin{equation}
    \Delta \sigma_n \,=\, b \cdot \left(\vphantom{\big()} \yieldf{\sigma_{n-7}}) - \yieldf[\target]{} \right)
    	\cdot \sigma_{n-7}
    \,=\, b \cdot \left(\vphantom{\big()} \inflaf{\sigma_{n-7}} - \yieldf[\target]{} \cdot \sigma_{n-7} \right)\;.
\end{equation}
The equilibrium points of the system, denoted by $\sigma^*$, are steady states, where the staking
rate no longer changes.
They are found by setting $\Delta \sigma_n = 0$, which yields the equilibrium condition
$\yieldf{\sigma^*} = \yieldf[\target]{}$.
To analyze the stability around such an equilibrium, we linearize the system.
Let $\sigma_n = \sigma^* + x_n$, where $x_n$ is a small perturbation.
By applying a first-order Taylor expansion to the model, we obtain a linear difference equation
that governs the evolution of the perturbation:
\begin{equation}
    x_n = x_{n-1} + b \cdot \left( \inflafp{\sigma^*} - \yieldf[\target]{} \right) \cdot x_{n-7}
\end{equation}
where $\inflafp{\sigma^*}$ is the derivative of the inflation function evaluated at the equilibrium point.
The corresponding characteristic
equation is derived by assuming that the solution to the
linear difference equation, $x_n = x_{n-1} + K_{\equi} \cdot x_{n-7}$, is of the
exponential form
$x_n = \lambda^n$.
This is a seventh-degree polynomial in $\lambda$, i.e. $\lambda^7 - \lambda^6 - K_{\equi} = 0$,
where we have defined the stability parameter $K_{\equi}$ as
$K_{\equi} = b(\inflafp{\sigma^*} - \yieldf[\target]{})$.
The equilibrium is locally stable if and only if all roots of this equation lie inside
the unit circle ($|\lambda| < 1$).
The stability boundary is reached when a root lies on the unit circle.
The analytical investigation in \Cref{sec:anal:math}
reveals that the system is stable for $K_{\equi}$ within the range:
\begin{equation}
    -2 \sin\left(\frac{\pi}{26}\right) < K_{eq} < 0
\end{equation}
which provides the approximate numerical stability condition $-0.241 < K_{eq} < 0$.
This criterion quantitatively links the stability of the system to the properties of the inflation
curve at its equilibrium points, particularly its slope, $\inflafp{\sigma^*}$.

\subsection{Analytical Investigation}\label{sec:anal:math}
The analytical investigation of the stability boundary for the characteristic equation $\lambda^7 - \lambda^6 - K_{eq} = 0$ proceeds by finding the values of the real parameter $K_{eq}$ for which at least one root has a modulus of one, i.e., $|\lambda|=1$. Such a root can be written as $\lambda = e^{i\omega}$. Substituting this into the equation gives:
\begin{equation*}
    e^{i7\omega} - e^{i6\omega} = K_{eq}
\end{equation*}
By applying Euler's formula, $e^{i\theta} = \cos(\theta) + i\sin(\theta)$, and separating the real and imaginary parts, we obtain two conditions:
\begin{align*}
    \text{Real part:} \quad & K_{eq} = \cos(7\omega) - \cos(6\omega) \\
    \text{Imaginary part:} \quad & \sin(7\omega) - \sin(6\omega) = 0.
\end{align*}
The second equation, $\sin(7\omega) = \sin(6\omega)$, yields two families of solutions for the frequency $\omega$. The first, $\omega = 2n\pi$, corresponds to a real root at $\lambda=1$. The second, $\omega = (2n+1)\pi/13$, corresponds to 13 different values on the unit circle. We find the critical values of $K_{eq}$ by evaluating the real part equation at these frequencies:
\begin{itemize}
    \item We check the real root $\lambda =  1$. This case  yields $1^7 - 1^6 - K_{eq} = 0$, which gives the stability boundary $K_{eq} = 0$. 
    \item For the other values $\omega_n=(2n+1)\pi /13 $ we can check that the maximum negative value of $\cos(7\omega_n)-cos(6\omega_n)$ is attained at $\omega = \pi/13$ (for $n=0$), giving:
    \begin{equation*}
        K_{eq} = \cos(7\pi/13) - \cos(6\pi/13)= -2\sin(\pi/26) \approx -0.241.
   \end{equation*}
\end{itemize}
By analyzing these boundary values, we find that the system remains stable when $-2\sin(\pi/26) < K_{eq} < 0$.

\subsection{Applications to the Stability Corridors}\label{sec:anal:obs}
We now apply the analytical stability criterion, $-0.241 < K_{eq} < 0$, to the three inflation
models, using a target yield of $\yieldf[\target]{} = 16.6\%$.
For the constant inflation model, the equilibrium occurs at $\sigma^* \approx 60.2\%$.
Since the derivative $\inflafp{\sigma^*}$ is zero, the stability parameter is
$K_{\equi} = -0.166 \cdot b$, and the system remains stable for a wide range of staker
sensitivity ($0 < b < 1.452$).
In contrast, for the \polkadot-style model, the equilibrium point at $\sigma^* \approx 51.6\%$
falls on the steep, exponentially decaying part of the inflation curve.
The large negative derivative, $\inflafp{\sigma^*} \approx -0.772$, yields a stability parameter
of $K_{\equi} = -0.938 \cdot b$.
Consequently, the system is stable only for a very narrow sensitivity range ($0 < b < 0.257$),
making it highly prone to oscillations.
Finally, for our proposed stability corridor model, the equilibrium at $\sigma^* \approx 21.3\%$ is
located on the first, gently sloped branch.
The resulting small positive derivative, $\inflafp{\sigma^*} = 0.025$, leads to
$K_{\equi} = -0.141 \cdot b$ and a very wide stability range ($0 < b < 1.709$), confirming that
this design is an even more robust against oscillatory behaviours.

	% To allow compilation of the file
% !TeX root = ./../main.tex

\section{Conclusions and Future Directions}\label{sec:conclusions}

In this paper, we have investigated the dynamic instability of staking rates in Proof-of-Stake
protocols.
We demonstrated, through a simple discrete-time model, studied by simulation and analytical
investigation that the interplay between yield sensitivity and
delayed staker responses is a powerful mechanism that can, by itself, generate significant
and persistent oscillations around the target equilibrium.

It is important to emphasize that this work is intended as a proof of concept.
The simulations presented are meant to qualitatively illustrate the stabilization effect of our
proposed dynamical-inflation function.
While the seven-day delay reflects the average lag observed by \citet{XXX:ConHeTan25} (and in any case it can vary from system to system) other model
parameters, such as the staker sensitivity $b$, have been chosen to reflect plausible agent behavior
but are not yet supported by systematic empirical studies.
Further research is necessary to refine these behavioural models and to statistically estimate
their parameters from on-chain data of specific decentralized economies,  allowing the dynamical-inflation function to adapt to the current economic state of the system.

In conclusion,  we believe that the stability corridor approach proposed and tested in this paper offers consistent advantages over traditional models. It eliminates excessive sensitivity by creating a yield plateau, nullifying the incentive  for stakers to react to minor fluctuations within the corridor, which in turn dampens oscillations.
Finally, the resulting predictability of returns reduces speculative behavior by lowering uncertainty  for stakers, thus attracting participants with a long-term orientation and simplifying protocol governance.

	\ \\
	\textbf{Acknowledgments}
	This paper has been written during the first research phase for the Palliora project
	(https://www.palliora.org), 
	which incorporates our staking approach for the token inflation.

	\printbibliography

\end{document}